# Characterization of the Autonomic Nervous System Activity in Females Classified According to Mood Scores During the Follicular Phase

Makiko Aoki*
Faculty of Health Science and Nursing
Juntendo University
Shizuoka, Japan
m.aoki.tt@juntendo.ac.jp

Mai Nishimura
Comprehensive Human Sciences
Research Group, Graduate School of
Comprehensive Human Sciences
University of Tsukuba
Ibaraki, Japan

Masato Suzuki
Graduate School of Science
University of Hyogo
Hyogo, Japan
suzuki@sci.u-hyogo.ac.jp
https://orcid.org/0000-0003-2040-3029

Eriko Terasawa
Comprehensive Human Sciences
Research Group, Graduate School of
Comprehensive Human Sciences
University of Tsukuba
Ibaraki, Japan

Hisayo Okayama
Faculty of Medicine
University of Tsukuba
Ibaraki, Japan
okayama@md.tsukuba.ac.jp
https://orcid.org/0000-0003-1253-9874

*Abstract*—Many sexually mature females suffer from premenstrual syndrome (PMS), but effective coping methods for PMS are limited due to the complexity of symptoms and unclear pathogenesis. Awareness has shown promise in alleviating PMS symptoms but faces challenges in long-term recording and consistency. Our research goal is to establish a convenient and simple method to make individual female aware of their own psychological, and autonomic conditions. In previous research, we demonstrated that participants could be classified into non-PMS and PMS groups based on mood scores obtained during the follicular phase. However, the properties of neurophysiological activity in the participants classified by mood scores have not been elucidated. This study aimed to classify participants based on their scores on a mood questionnaire during the follicular phase and to evaluate their autonomic nervous system (ANS) activity using a simple device that measures pulse waves from the earlobe. Participants were grouped into Cluster I (high positive mood) and Cluster II (low mood). Cluster II participants showed reduced parasympathetic nervous system activity from the follicular to the menstrual phase, indicating potential PMS symptoms. The study demonstrates the feasibility of using mood scores to classify individuals into PMS and non-PMS groups and monitor ANS changes across menstrual phases. Despite limitations such as sample size and device variability, the findings highlight a promising avenue for convenient PMS self-monitoring.

## I. Introduction

Many sexually mature females exhibit a variety of psychological, autonomic, and physiological symptoms, such as irritability, mood swings, hot flashes, palpitations, anorexia, loss of appetite, abdominal pain, and breast tenderness, which occur periodically during the menstrual cycle. When these symptoms are particularly intense during the premenstrual (luteal) phase, premenstrual syndrome (PMS) is diagnosed [1]. The gynecologist diagnoses PMS based on a diary record of the patient's symptoms over several months. The difficulty of keeping a long-term record makes it challenging to assess PMS accurately. Therefore, the prevalence of PMS is highly variable, ranging from 12% (study in France) to 98% (study in Iran) in sexually mature females, depending on the type of research [2]. Since PMS symptoms are complex and the mechanisms of onset remain unclear, effective coping strategies are limited. Awareness and acceptance of PMS have been reported to be associated with an increased perceived ability to cope with PMS [3], offering a defensive effect against premenstrual symptoms [4]. Awareness and acceptance through habitual mindfulness have been reported to alleviate PMS symptoms [5]. Moreover, the use of menstrual tracking smartphone applications has been reported to significantly reduce the incidence rates of depression and dysmenorrhea [6]. Thus, awareness and acceptance of one's state can potentially cope with PMS symptoms. However, this approach necessitates diary-like records over a long period (covering three menstrual cycles), posing a challenge to maintaining consistent records.

For a simple and rapid assessment of PMS status, it is effective to use a questionnaire that describes symptoms on the spot and in retrospect of past menstrual cycles, rather than a daily questionnaire such as the Daily Record of Severity of Problems (DRSP) [7], [8]. The Menstrual Distress Questionnaire (MDQ) is a questionnaire that reflects on the past menstrual cycle and describes the psychological state of each cycle and has been used to identify symptoms related to menstruation [9]. However, it has been revealed that MDQ is unsuitable for identifying PMS [10]. Although there are reported scales for assessing PMDD [11], which is considered a severe form of PMS, validated scales for on-the-spot evaluation of PMS are extremely limited [12], [13]. The profile of Mood States 2nd edition (POMS2) is a scale for evaluating moods [14], and it has been reported that PMS participants exhibit characteristic POMS scores during each menstrual cycle [15], [16]. Additionally, a significant decrease in the positive mood sub-scale (Vigor-Activity: VA and Friendliness: F) of POMS2 during the follicular phase in the PMS patients compared to the non-PMS patients has been reported. There was also an increase in sub-scales indicating negative mood, thereby demonstrating the ability to distinguish between PMS and non-PMS patients based on the score patterns of POMS2 sub-scales [17]. However, the reproducibility of this classification method has not been verified due to the small number of participants.

*Corresponding Author
This work was supported by JSPS KAKENHI Grant Numbers 21K12794 to HO and 22K12950 to MA.

On the other hand, to understand the pathogenesis of PMS and to objectively diagnose PMS symptoms, research has been conducted to measure biosignals such as electrocardiograms and brain functions in participants with PMS and PMDD and to explore characteristic biosignals of these participants. Functional brain imaging techniques such as functional magnetic resonance imaging (fMRI) [18] and electroencephalogram (EEG) [19] have revealed the dysfunction in the amygdala and prefrontal cortex (PFC) regions responsible for simple mental arithmetic tasks and emotional processing. Using near-infrared spectroscopy (NIRS), which can easily measure blood flow optically, our study also found that blood flow in the PFC of female students in their 20s with PMS symptoms during a two-back task, one of the cognitive tasks, was reduced compared to non-PMS participants [16]. Additionally, dysfunction in the amygdala region has been associated with the hypothalamus, which regulates autonomic nervous system (ANS) activity. A reduction in parasympathetic nervous activity in PMS participants during the luteal and menstrual phases has been revealed through power spectrum analysis of electrocardiograms (ECGs) [20]. The measurement of ANS activity is expected to be applied to the awareness of PMS symptoms, as it requires no task presentation and can be measured in a short time.

Based on the above background, the combination of a simple questionnaire and daily ANS activity measurement can be used to achieve convenient and accurate awareness of one's mental condition. In this study, the POMS2 subscale of the follicular phase was used to classify the participants into PMS and non-PMS groups, to understand the characteristics of ANS activity during the follicular and menstrual phases of each group. In anticipation of achieving future convenient and casual awareness, ANS activity was assessed by power spectral analysis of pulse waves obtained from the earlobe by photoplethysmography (PPG).

## II. EXPERIMENTAL METHODS

### A. Requirements for participants in this study

Participants consisted of undergraduate and graduate female students in their early 20s. The following information was collected for each participant: age, height, weight, body mass index (BMI), which is obtained from height and weight, age at menarche, daily medication use, menstrual cycle length, duration of menstruation, duration of the last two menstrual periods, menstruation-related medication uses and drug names, and awareness of menstrual symptoms. The awareness of menstrual symptoms was assessed using the MDQ. The MDQ comprises 47 questions and is used to evaluate physical, mood, and arousal symptoms before, during, and after menstruation. The A-type MDQ, which requires participants to recall and describe symptoms at three periods: premenstrual, during menstruation, and postmenstrual, was utilized. Participants were selected based on the following criteria:

- Menstrual cycles range from 25 to 38 days.
- BMI between 18.5 and 25.0.
- Not using medications that affect autonomic nerve system balance.
- No smoking habit.
- Not taking oral contraceptives within 6 months.
- No experience of pregnancy or childbirth.

### B. Experimental Procedure

Participants underwent measurements of their blood pressure (HEM-FM31, OMRON Corp., Kyoto, Japan), mood (shortened Japanese version of POMS2), and autonomic nervous activity measured during the menstrual and follicular phases. The day on which menstruation occurred was defined as day 1 menstruation, and the menstrual phase was designated as day 2 of menstruation. The follicular phase was defined as 5-7 days after the menstrual phase. ANS activity was evaluated based on heart rate variability obtained using an earlobe-mounted PPG (Inner Balance Scan, Biocom Technologies, Poulsbo, WA). After wearing the instrument, participants remained seated in a quiet state for 5 minutes. Subsequently, the plethysmogram was measured for 5 minutes. The Fast Fourier transform method was applied to the obtained plethysmograph, and the power spectra were calculated. Power spectra in the frequency range of 0.004-0.15 Hz were designated as low-frequency components (LF), while those in the frequency range of 0.15-0.4 Hz were designated as high-frequency components (HF).

### C. Data Analysis

Participants were classified into two groups using Ward's method based on their scores of POMS2 during the follicular phase. The obtained ANS indices (heart rate (HR), R-R interval (RRI), LF, and HF) were evaluated for outliers using the Grubbs outlier test. When comparing values of MDQ, POMS, HR, RRI, and LF/HF between the two groups, the Mann-Whitney U-test was employed.

### D. Ethical Consideration

This study was conducted after obtaining approval from the Medical Ethics Committee of Medicine and Health Sciences in the University of Tsukuba Faculty of (Approval No. 1650-1).

## III. RESULTS AND DISCUSSION

### A. Classification of Participants According to POMS2 Scores

ANS activity during the follicular and menstrual phases was obtained from 19 participants, and as a result of the Grubbs test, data from 15 participants during the menstrual phase and 17 participants during the follicular phase were used for analysis. Participants were classified into two groups, Cluster I and Cluster II, using Ward's method based on their POMS2 scores during the follicular phase. The classification results using Ward's method are shown in Fig. 1A. Cluster I consisted of 11 participants, while Cluster II comprised 6 participants. The basic properties, blood pressure, and respiratory rate of all participants and those classified into Cluster I and Cluster II are shown in Table 1. No significant differences were observed in any of the items, indicating similar basic properties across all groups. The distribution of scores for the subscales of POMS2 obtained during the follicular phase in Cluster I and Cluster II is shown in Fig 1B. Participants in Cluster I exhibited significantly higher scores in parameters for positive mood (F and VA), compared to

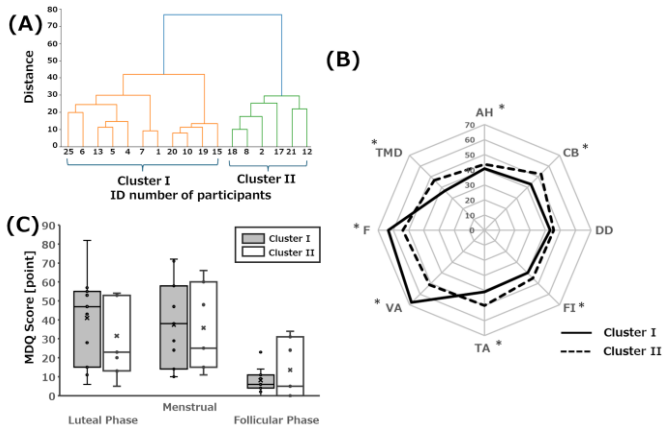

Fig. 1. Classification of participants into two groups according to the POMS2 sub-scale scores during the follicular phase. (A) The two Clusters classified using Ward's method and the ID numbers of the participants belonging to each Cluster. (B) Distribution of the mean values of each sub-scale of POMS2 obtained from the participants in Cluster I and Cluster II, respectively. (C) MDQ scores obtained from participants in Clusters I and Cluster II, respectively. * $p < 0.05$.

Cluster II ($p<0.05$). In addition, the negative mood subscales, including anger-hostility (AH), confusion-bewilderment (CB), depression-dejection (DD), fatigue-inertia (FI), tension-anxiety (TA), and total mood disturbance (TMD), were lower in Cluster I, with AH, CB, FI, TA, and TMD being significantly lower ($p<0.05$). This distribution of scores on the POMS2 sub-scales during the follicular phase closely resembled the results of previous our study conducted on female university students in their 20s with PMS symptoms [17]. In this previous study, participants classified into Cluster I, with significantly higher F and VA scores, were categorized by gynecologists as non-PMS participants based on PMDD scale scores. Conversely, Cluster II included PMS participants, PMDD participants, or participants who appeared to have mild depressive symptoms according to their scores on the self-rating depression (SDS), which is an index for assessing depression. Therefore, participants classified in Cluster I in this study were considered to be non-PMSs, while those classified into Cluster II were considered to have PMS or PMDD symptoms. It is interesting to note that the participants in the present study are a completely different population from the participants in our previous study, with only the same basic properties (age, menstrual cycle, etc.). In addition, the location of the current study was different from our previous study. Despite these differences, adopting Ward's method to the scores in the sub-scales of POMS2 yields two distinct groups: one with significantly higher scores in positive mood (Cluster I) and another with higher scores in negative mood (Cluster II). The present distribution of the scores on the sub-scales of the POMS2 for these two groups closely resembled that obtained in previous research. These results demonstrate that the classification of participants based on POMS 2 sub-scale scores during the follicular phase is a highly robust and effective method for easily distinguishing between non-PMS and PMS/PMDD groups.

In the Cluster I group, the MDQ scores during premenstrual (corresponding to the late luteal phase) and menstrual phases were significantly higher compared to those during the follicular phase (Fig. 1C). A higher MDQ score indicates greater physical and psychological symptoms. Therefore, participants in Cluster I likely experienced physical and psychological symptoms during the premenstrual and menstrual phases, and these symptoms were likely relieved after menstruation. In a previous study of healthy Japanese female university students, the average MDQ scores during the postmenstrual (follicular phase) were significantly higher than those in the premenstrual (luteal phase) [21]. The MDQ scores obtained in the present study showed a similar trend to the previous study, suggesting the validity of the MDQ scores obtained in the present study. Cluster II showed a similar distribution of premenstrual, menstrual, and postmenstrual MDQ scores. The average value of postmenstrual MDQ scores in Cluster II was higher than that in Cluster I, suggesting that Cluster II may continue to experience physical and mental discomfort even during the postmenstrual phase. There were no significant differences between Cluster I and Cluster II in MDQ values during any phases. While the MDQ has a good record of quantifying the intensity of symptoms in each menstrual cycle, issues of accuracy in assessing PMS have been reported [10]. The lack of differences in MDQ scores between the Cluster I and Cluster II groups suggests the difficulty in identifying PMS by MDQ scores.

*B. Evaluation of ANS Activity*

The ANS activity properties of the participants classified as Cluster I and Cluster II were compared (Fig. 2). HR and RRI, directly obtained from pulse wave, did not differ significantly between Cluster I and Cluster II during either the follicular or menstrual phases. The LF and HF, which are used as indicators of ANS activity, were compared between Cluster I and Cluster II. LF slightly decreased during the menstrual phase compared to the follicular phase in both Cluster I and Cluster II, which implies a decrease in parasympathetic nerve system activity. However, there were no statistically significant differences in LF values in either group, suggesting no change in LF values in either Cluster I or Cluster II. Similarly, HF decreased during the menstrual phase compared to the follicular phase in both Cluster I and Cluster II. Particularly in Cluster II, HF significantly decreased during the menstrual phase compared to the follicular phase. The p-value of HF between the follicular and menstrual phases in Cluster II was 0.11, indicating a trend of decreased HF values during the menstrual phase. Since HF reflects parasympathetic ANS, this result suggests a trend toward decreased parasympathetic nervous system activity in Cluster II participants from the follicular to the menstrual phase. A previous study reported a significant decrease in HF during the luteal phase (premenstrual) in the PMS group compared to HF values during the follicular phase [22]. The substantial decrease in HF values from the follicular phase to the menstrual phase in the Cluster II group likely reflects a reduction in parasympathetic activity associated with PMS symptoms. These results suggest that even with a simple earlobe-mounted PGG device, the reduction in parasympathetic activity specific to PMS participants from the premenstrual to the menstrual period can be detected.

TABLE I. BASIC PROPERTIES OF ALL PARTICIPANTS, PARTICIPANTS IN CLUSTER I, AND PARTICIPANTS IN CLUSTER II

|  | Participants | | |
| --- | --- | --- | --- |
|  | *All* | *Cluster I* | *Cluster II* |
| Menstrual Cycle, day | 32.63 | 33.00 | 32.00 |
| Menstrual Period, day | 6.05 | 6.27 | 5.57 |
| Systolic Blood Pressure (FP), mmHg | 107.74 | 109.27 | 105.57 |
| Diastolic blood pressure (FP), mmHg | 72.26 | 73.64 | 70.86 |
| Respiration rate (FP), min$^{-1}$ | 18.06 | 17.73 | 18.67 |
| Systolic Blood Pressure (menstruation), mmHg | 106.68 | 107.45 | 106.57 |
| Diastolic blood pressure (menstruation), mmHg | 71.37 | 71.09 | 72.71 |
| Respiration rate (menstruation), min$^{-1}$ | 17.67 | 16.60 | 19.80 |

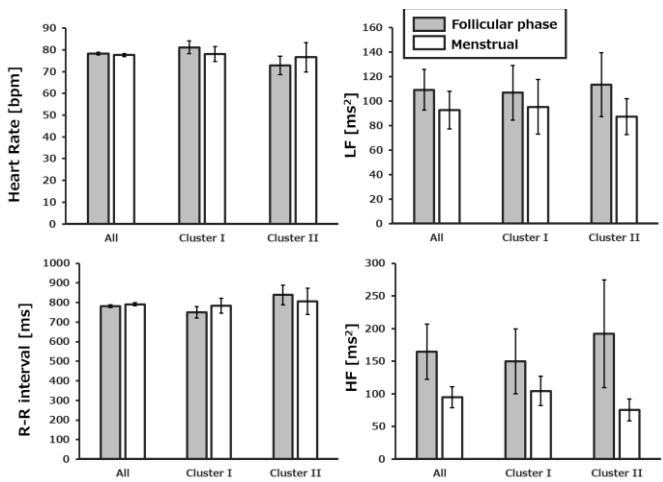

Fig. 2. Mean values of ANS activity parameters (HR, RRI, LF, and HF) obtained from participants in Cluster I and Cluster II, respectively.

## IV. Conclusion

In this study, participants were classified based on the POMS2 sub-scale scores during the follicular phase, aiming to characterize their ANS activity. As a result of classifying participants into two groups, they were categorized into a group with a high positive mood (Cluster I) and a group with a low mood (Cluster II). The distribution of POMS2 sub-scale scores in these two groups was comparable to that obtained in our previous study. These findings suggest that the POMS2 subscale scores can classify participants into two groups: those with a high likelihood of PMS and those without, even if the timing and institution of implementation differ. Furthermore, it was revealed that participants in Cluster II, considered to be PMS symptoms, showed a decreasing trend in HF values, indicating parasympathetic nervous activity, from the follicular phase to the menstrual phase. These results suggest the potential for detecting PMS through mood assessment during the follicular phase and measuring pulse waves from the luteal phase to the menstrual phase. However, this study had a small sample size of 19 participants. Additionally, since measurements of PPG were conducted using a simple earlobe-mounted device, there was a large variability in the measured values. Therefore, it is necessary to increase the number of participants and simultaneously measure both pulse wave-based simple heart rate variability measurement devices and electrocardiograms to validate the measurement values.

## Acknowledgment

M. A., M. N., and E. T. performed experiments. M. A. and M. S. analyzed experimental data. M. N. and H. O. conceived the project. M. A. and M. S. wrote the paper.